\documentclass[twocolumn]{article}
\usepackage[preprint]{spconfa4}
\usepackage{amsmath,graphicx}

\usepackage{bm}
\usepackage{amssymb}
\usepackage{amsfonts}
\usepackage{color}
\usepackage{cite}
\usepackage{booktabs}
\usepackage{subfig}

\renewcommand{\paragraph}[1]{\noindent\quad#1\\}

\title{
Physics-informed convolutional neural network with bicubic spline interpolation for sound field estimation
}
\name{Kazuhide Shigemi, Shoichi Koyama, Tomohiko Nakamura, Hiroshi Saruwatari}
\address{Graduate School of Information Science and Technology, The University of Tokyo\\7-3-1 Hongo, Bunkyo-ku, Tokyo 113-0033, Japan}
\begin{document}
\ninept
\maketitle
\begin{abstract}
A sound field estimation method based on a physics-informed convolutional neural network (PICNN) using spline interpolation is proposed. Most of the sound field estimation methods are based on wavefunction expansion, making the estimated function satisfy the Helmholtz equation. However, these methods rely only on physical properties; thus, they suffer from a significant deterioration of accuracy when the number of measurements is small. Recent learning-based methods based on neural networks have advantages in estimating from sparse measurements when training data are available. However, since physical properties are not taken into consideration, the estimated function can be a physically infeasible solution. We propose the application of PICNN to the sound field estimation problem by using a loss function that penalizes deviation from the Helmholtz equation. Since the output of CNN is a spatially discretized pressure distribution, it is difficult to directly evaluate the Helmholtz-equation loss function. Therefore, we incorporate bicubic spline interpolation in the PICNN framework. Experimental results indicated that accurate and physically feasible estimation from sparse measurements can be achieved with the proposed method. 

\end{abstract}
\begin{keywords}
sound field estimation,
physics-informed neural networks,
Helmholtz equation,
spline interpolation
\end{keywords}

\section{Introduction}
\label{sec:intro}

Sound field estimation (or reconstruction) is aimed at estimating or interpolating an acoustic field from discrete measurements of sensors (microphones). Such estimation has a broad range of applications, such as spatial audio recording for virtual/augmented reality~\cite{poletti2005three-dimensional,Koyama:IEEE_J_ASLP2013,Iijima:JASA_J_2021}, visualization of sound fields~\cite{maynard1985nearfield}, identification of sources~\cite{teutsch2006acoustic}, and spatial active noise control~\cite{8264817,Ito:ICASSP2019,Koyama:IEEE_ACM_J_ASLP2021}; therefore, a large number of sound field estimation methods have been studied. 

A typical approach to sound field estimation is based on wavefunction expansion~\cite{williams1999fourier}, which fairly approximates any solution of the homogeneous Helmholtz equation. Representing an acoustic field by a combination of plane wave functions or spherical wavefunctions allows us to interpolate the sound field in a source-free region by estimating expansion coefficients~\cite{poletti2005three-dimensional,Koyama:IEEE_J_ASLP2013,Samarasinghe2014Wavefield}. The kernel interpolation for the sound field~\cite{Ueno:IEEE_SPL2018,Ueno:IEEE_J_SP_2021}, which corresponds to the infinite-dimensional wavefunction expansion, constrains the interpolated function to satisfying the Helmholtz equation. Thus, current sound field estimation methods take advantage of physical properties of the acoustic field. However, these methods basically do not adapt to the acoustic environment in which the estimation is performed. Therefore, the estimation accuracy deteriorates greatly when only a small number of microphones are available.

Learning the properties of the acoustic environment will be a promising strategy to increase the estimation accuracy when the available number of microphones is limited but training data are available. Recent studies have shown that sound field estimation methods based on deep neural networks (DNNs) make interpolation from sparse measurements possible by learning from data~\cite{lluis2020sound,strom2021deep}. However, these methods do not exploit physical properties and are largely dependent on training data. Thus, the estimated sound field can be a physically infeasible solution. 


Several studies on incorporating physical properties into a DNN framework have been conducted in recent years. For example, a physics-informed neural network (PINN)~\cite{raissi2019physics,karniadakis2021physics} learns a function that returns a value corresponding to positional inputs using a loss function penalizing deviations from the governing equation, making the network have desired physical properties. A similar idea is applied to a convolutional-neural-network (CNN) framework in the physics-informed CNN (PICNN)~\cite{zhao2021physicsinformed} to utilize various spatial resolution features of input images. The effectiveness of these methods is mainly validated in a forward problem, e.g., physical simulation. 


We propose a learning-based sound field estimation method exploiting physical properties. We apply the PICNN framework to the sound field estimation, which is an inverse problem to estimate a distribution satisfying the Helmholtz equation from a sparse set of measurements. We define the Helmholtz-equation loss function that measures the deviation of the CNN output from functions satisfying the Helmholtz equation. However, the CNN output is a spatially discretized acoustic pressure distribution. Since the Helmholtz equation involves a Laplacian operator, it is difficult to evaluate the Helmholtz-equation loss function from the CNN output. Therefore, we incorporate the bicubic spline interpolation in the PICNN-based sound field estimation. Numerical experiments are conducted to evaluate the proposed method, and compare it with kernel-interpolation-based and CNN-based methods.


\section{Problem statement and prior works}
\subsection{Sound field estimation problem}\label{subsec:problem}
Suppose that a target region $\Omega$ is an open subset of $\mathbb{R}^{3 \ \text{or} \ 2}$. The acoustic pressure at the position $\bm{r}\in\Omega$ and angular frequency $\omega\in\mathbb{R}$ is denoted by $u(\bm{r},\omega)$ ($u: \Omega \times \mathbb{R} \to \mathbb{C}$). We assume that the region $\Omega$ does not contain sources and the acoustic field is stationary. Then, we set $N$ evaluation points inside $\Omega$, whose index set and positions are denoted by $\mathcal{N}$ and $\{\bm{r}_n\}_{n\in\mathcal{N}}$ ($|\mathcal{N}|=N$), respectively. $M$ observations $\{s_m\}_{m\in\mathcal{M}}$ are taken from a subset of the evaluation points $\mathcal{M}\subseteq\mathcal{N}$ ($|\mathcal{M}|=M$). These observations are obtained by using pressure microphones at the positions $\{\bm{r}_m\}_{m\in\mathcal{M}}$. Our goal is to estimate $\{u(\bm{r}_n,\omega)\}_{n\in\mathcal{N}} \subset \mathbb{C}^N$ from $\{s_m\}_{m\in\mathcal{M}} \subset \mathbb{C}^M$. Hereafter, $\omega$ is omitted for notational simplicity. 

\subsection{Prior works on sound field estimation}\label{subsec:kernel}
There exist several well-established methods for the sound field estimation problem~\cite{poletti2005three-dimensional,Koyama:IEEE_J_ASLP2013}. The method based on kernel ridge regression proposed in \cite{Ueno:IWAENC2018} enables us to interpolate a pressure field from a discrete set of observations using a kernel function that guarantees that the estimated function satisfies the Helmholtz equation. This method corresponds to the estimation based on infinite-dimensional expansion into plane wave functions or spherical wavefunctions~\cite{Ueno:IEEE_SPL2018,Ueno:IEEE_J_SP_2021}. However, this kernel interpolation method does not adapt to the acoustic environment in which the estimation is performed, with a few exceptions~\cite{Horiuchi:WASPAA2021}, because the kernel function is generally defined in analytical form. In the subsequent sections, we consider sound field estimation methods that learn or adapt to the features of the acoustic environment.

\section{CNN-based sound field estimation method}\label{sec:baseline}
First, we introduce a CNN-based sound field estimation method for the problem defined in Sect.~\ref{subsec:problem} that incorporates the features learned from data into the estimation. This method is a simplification of the method proposed in \cite{lluis2020sound,strom2021deep} for comparison; therefore, we refer to this CNN-based method as the baseline method. 

To simplify the problem, the target region $\Omega$ is supposed to be a 2D rectangular region on the $xy$-plane with its center at the origin in a 2D space.
The evaluation points are regularly arranged on $\Omega$, whose index $n$ is redefined by $i$ and $j$ for each side of the rectangle. 
The number of evaluation points on each side is $I$ or $J$, i.e., $IJ=N$.
The points at the corners, i.e., $(i,j)=(1,1)$, $(1,J)$, $(I,1)$, and $(I,J)$, are on the corners of $\Omega$. 
Now, the problem to be solved is to estimate $\{u(\bm{r}_{i,j})\}_{(i,j)\in\mathcal{N}}$ from the observations $\{s_m\}_{m\in\mathcal{M}}$.

In the baseline method, a CNN receives an input tensor $\bm{S}\in\mathbb{R}^{3\times I\times J}$ and returns an output tensor $\hat{\bm{U}}\in \mathbb{R}^{2\times I\times J}$. The input tensor $\bm{S}$ has three channels, where the first two channels correspond to the real and imaginary parts of the observations, and the last channel is the mask matrix with 1 for the elements of the observation points and 0 otherwise. The output tensor $\hat{\bm{U}}$ has two channels, where these channels correspond to the real and imaginary parts of the pressure at the evaluation points.


The loss function $L_{\mathrm{D}}$ is defined as the mean square error between the estimated pressure $\{\hat{u}(\bm{r}_{i,j})\}_{(i,j)\in\mathcal{N}}$, which corresponds to the elements of the output $\hat{\bm{U}}$, and the ground truth $\{u(\bm{r}_{i,j})\}_{(i,j)\in\mathcal{N}}$ as
\begin{equation}\label{eq:dataLoss}
 L_{\mathrm{D}}:=\frac{1}{N}\sum_{(i,j)\in \mathcal{N}} | u(\bm{r}_{i,j})-\hat{u}(\bm{r}_{i,j}) |^2.
\end{equation}
$L_{\mathrm{D}}$ is called the \textit{data loss function}. In the baseline method, $L_{\mathrm{D}}$ is used as the total loss function $L$, i.e., $L=L_{\mathrm{D}}$.
The network is trained to minimize this loss function $L$ by using a set of training data, i.e., the input and output tensors consisting of $\{u(\bm{r}_{i,j})\}_{(i,j)\in\mathcal{N}}$. 




\section{Proposed method}\label{sec:proposed}
The baseline method in Sect.~\ref{sec:baseline} is one of the simple approaches to learning-based sound field estimation. However, there is no guarantee that the estimated pressure field is a physically feasible solution because CNN is simply designed to generate an output that minimizes the data loss function.

We propose a method that incorporates physical properties into the CNN-based sound field estimation method. Since the estimate $\hat{u}(\bm{r})$ is a pressure field, $\hat{u}(\bm{r})$ should satisfy the Helmholtz equation:
\begin{equation}\label{eq:HHeq}
    \left(\Delta+k^2\right)\hat{u}(\bm{r})=0,
\end{equation}
where $\Delta$ is the Laplacian and $k$ is the wavenumber. However, it is difficult to exactly restrict the estimate $\hat{u}(\bm{r})$ to a member of the solution space of \eqref{eq:HHeq} in the CNN framework. Hence, we consider to relax the constraint by formulating the \textit{Helmholtz-equation loss function} $L_{\mathrm{H}}$ that penalizes the deviation of the estimate from the Helmholtz equation. 



To define $L_{\mathrm{H}}$, it is necessary to calculate \eqref{eq:HHeq} for the output of CNN. However, the CNN output $\hat{\bm{U}}$ is a spatially discretized pressure at $\{\bm{r}_{i,j}\}_{(i,j)\in\mathcal{N}}$.
That means \eqref{eq:HHeq} is not computable because $\hat{u}$ is a continuous function of $\bm{r}\in\Omega$.
Furthermore, the difference approximation of \eqref{eq:HHeq} is not preferable, especially when the intervals of the evaluation points are not sufficiently small. Therefore, we consider interpolating the CNN output to obtain $\hat{u}$ for $\bm{r}\in\Omega$ from $\hat{\bm{U}}$. Then, $L_{\mathrm{H}}$ becomes computable by using the interpolated function. 




\subsection{Bicubic spline interpolation}\label{subsec:interpolation}

There exist numerous interpolation methods; however, in this scenario, the 2D function must be interpolated from discrete values. Moreover, the interpolated function must be twice differentiable, i.e., in $\mathcal{C}^2$, and its twice differential values should generally be non-zero. We apply bicubic spline interpolation~\cite{helmuth1993twodim} that meets the requirements described above. 


The real and imaginary parts of $\hat{u}(x,y)$, defined by $\hat{u}^{\mathrm{Re}}(x,y)$ and $\hat{u}^{\mathrm{Im}}(x,y)$, respectively, are interpolated separately. Both $\hat{u}^{\mathrm{Re}}(x,y)$ and $\hat{u}^{\mathrm{Im}}(x,y)$ are divided into $(I-1)\times(J-1)$ small rectangular patches $\{\Omega_{i,j}\}_{(i,j)\in\bar{\mathcal{N}}}$ where the $(i,j)$th patch is in $[x_{i},x_{i+1}] \times [y_{j},y_{j+1}]$ with $(x_{i},y_{j}):=\bm{r}_{i,j}$, and $\bar{\mathcal{N}}$ is the index set excluding the points on the two sides of $\Omega$, $ \{(i,J)\}_{i=1}^I \cup \{(I,j)\}_{j=1}^J$. In the bicubic spline interpolation, the functions in each patch $h_{i,j}^{\mathrm{Re}}(x,y)$ and $h_{i,j}^{\mathrm{Im}}(x,y)$ are approximated by 2D cubic functions.
Therefore, for $p\in\{\mathrm{Re},\mathrm{Im}\}$, $h_{i,j}^{p}(x,y)$ is obtained as
\begin{align}\label{eq:iterpolated}
    h_{i,j}^{p}(x,y) &= \bm{g}\left(x-x_{i}\right)^\mathsf{T} \bm{A}_{i,j}^{p}\bm{g}\left(y-y_{j}\right),
\end{align}
where $\bm{A}_{i,j}^{p}\in\mathbb{R}^{4\times 4}$ and $\bm{g}(z):=[1,z,z^2,z^3]^\mathsf{T}$. The elements of $\{\bm{A}_{i,j}^{p}\}_{(i,j)\in\bar{\mathcal{N}}}$ are calculated by sequentially solving linear equations using the following.
\begin{align*}
    \begin{cases}
    \displaystyle \hat{u}^{p}(x_i,y_j) & \text{for} \quad (i,j)\in\mathcal{N}\\
    \hat{u}_x^{p}:= \dfrac{\partial \hat{u}^{p}}{\partial x} & \text{for} \quad (i,j)\in\mathcal{N}_x \\[5pt]
    \displaystyle \hat{u}_y^{p} := \frac{\partial  \displaystyle \hat{u}^{p}}{\partial y} & \text{for} \quad (i,j)\in\mathcal{N}_y \\[5pt]
    \displaystyle \hat{u}_{x,y}^{p}:=\frac{\partial^2  \displaystyle \hat{u}^{p}}{\partial x \partial y} & \text{for} \quad (i,j)\in\mathcal{N}_{x,y}
    \end{cases}
\end{align*}
Here, $\mathcal{N}_x:=\{(1,j)\}_{j=1}^J\cup\{(I,j)\}_{j=1}^J$, $\mathcal{N}_y:=\{(i,1)\}_{i=1}^I\cup\{(i,J)\}_{i=1}^I$, and $\mathcal{N}_{x,y}:=\{(1,1),(I,1),(1,J),(I,J)\}$.
Then, $\hat{u}^{\mathrm{Re}}(x,y)$ and $\hat{u}^{\mathrm{Im}}(x,y)$ are approximated as a concatenation of the patches $h_{i,j}^{\mathrm{Re}}(x,y)$ and $h_{i,j}^{\mathrm{Im}}(x,y)$, respectively. Thus, $\hat{u}(x,y)$ is obtained as $\hat{u}^{\mathrm{Re}}(x,y) + \mathrm{j} \hat{u}^{\mathrm{Im}}(x,y)$.

\subsection{Computation of Helmholtz-equation loss function}
To compute the Helmholtz-equation loss function, the partial derivatives of $\hat{u}$ are added to the CNN output. Specifically, the output tensor $\hat{\bm{U}}\in \mathbb{R}^{(4\times 2)\times I\times J}$ consists of $\{\hat{u}^{p}(x_i,y_j)\}_{(i,j)\in\mathcal{N}}$, $\{\hat{u}^{p}_x(x_i,y_j)\}_{(i,j)\in\mathcal{N}_x}$, $\{\hat{u}^{p}_y(x_i,y_j)\}_{(i,j)\in\mathcal{N}_y}$, and $\{\hat{u}^{p}_{xy}(x_i,y_j)\}_{(i,j)\in\mathcal{N}_{x,y}}$, i.e., four channels for each real and imaginary part. Using the output $\hat{\bm{U}}$ and the procedure shown in Sect.~\ref{subsec:interpolation}, we can obtain the continuous $\hat{u}(x,y)$.

The Helmholtz-equation loss function $L_{\mathrm{H}}$ is defined as
\begin{align}\label{eq:HELossdef}
    L_{\mathrm{H}}&:=\frac{1}{S_\Omega}\int_{\Omega}\left|\left(\Delta +k^2\right)\hat{u}(x,y)\right|^2\mathrm{d}x\mathrm{d}y,
\end{align}
where $S_\Omega$ represents the area of $\Omega$.
This function $L_{\mathrm{H}}$ is used to evaluate the deviation of $\hat{u}$ from the function that satisfies the Helmholtz equation.
On the basis of the bicubic spline interpolation,  the integration in \eqref{eq:HELossdef} is computed as
\begin{align}\label{eq:HELosstrans}
    & \int_{\Omega}\left|\left(\Delta +k^2\right)\hat{u}(x,y)\right|^2\mathrm{d}x\mathrm{d}y \notag\\
    &=\sum_{(i,j)\in\bar{\mathcal{N}}}\int_{\Omega_{i,j}}\left|\left(\Delta +k^2\right)\hat{u}(x,y)\right|^2\mathrm{d}x\mathrm{d}y. \notag\\
    &\simeq\sum_{(i,j)\in\bar{\mathcal{N}}}\left[\int_{\Omega_{i,j}}\left|\left(\Delta +k^2\right)\bm{g}\left(x-x_{i}\right)^\mathsf{T} \bm{A}^{\mathrm{Re}}_{i,j}\bm{g}\left(y-y_{j}\right)\right|^2\mathrm{d}x\mathrm{d}y\right.\nonumber \\
    & \quad\quad\quad +\left.\int_{\Omega_{i,j}}\left|\left(\Delta +k^2\right)\bm{g}\left(x-x_{i}\right)^\mathsf{T} \bm{A}^{\mathrm{Im}}_{i,j}\bm{g}\left(y-y_{j}\right)\right|^2\mathrm{d}x\mathrm{d}y\right].
\end{align}
Here, each patch $\Omega_{i,j}$ is assumed to be a square of the length $l$ for simplicity. Then, the integration in \eqref{eq:HELosstrans} is analytically derived as
\begin{align}\label{eq:HELosstrans_final}
    &\int_{\Omega_{i,j}}\left|\left(\Delta +k^2\right)\bm{g}\left(x-x_{i}\right)^\mathsf{T} \bm{A}^{p}_{i,j}\bm{g}\left(y-y_{j}\right)\right|^2\mathrm{d}x\mathrm{d}y \notag\\ 
    &=\mathrm{sum}\left[\bm{A}^{p}_{i,j} \bm{C}_1 \left(\bm{A}^{p}_{i,j}\right)^\mathsf{T}\otimes \bm{C}_2\right.\notag\\
    &\quad\quad\quad+\bm{A}^{p}_{i,j} \bm{C}_2 \left(\bm{A}^{p}_{i,j}\right)^\mathsf{T}\otimes \bm{C}_1\notag\\ 
    &\quad\quad\quad+k^4\bm{A}^{p}_{i,j} \bm{C}_1 \left(\bm{A}^{p}_{i,j}\right)^\mathsf{T}\otimes \bm{C}_1\notag\\
    &\quad\quad\quad+2\bm{A}^{p}_{i,j} \bm{C}_3^\mathsf{T} \left(\bm{A}^{p}_{i,j}\right)^\mathsf{T}\otimes \bm{C}_3\notag\\
    &\quad\quad\quad+2k^2\bm{A}^{p}_{i,j} \bm{C}_3 \left(\bm{A}^{p}_{i,j}\right)^\mathsf{T}\otimes \bm{C}_1 \notag\\
    &\quad\quad\quad\left.+2k^2\bm{A}^{p}_{i,j} \bm{C}_1 \left(\bm{A}^{p}_{i,j}\right)^\mathsf{T}\otimes \bm{C}_3^\mathsf{T}\right],
\end{align}
where $\mathrm{sum}[\cdot]$ is a function that returns the sum of all elements of an input matrix, $\otimes$ means the Kronecker product, and $\bm{C}_1$, $\bm{C}_2$, and $\bm{C}_3$ are matrices defined as follows:
\begin{align}\label{eq:Cdef}
    \bm{C}_1&=
    \begin{pmatrix}
        l & l^2/2 & l^3/3 & l^4/4 \\
        l^2/2 & l^3/3 & l^4/4 & l^5/5 \\
        l^3/3 & l^4/4 & l^5/5 & l^6/6 \\
        l^4/4 & l^5/5 & l^6/6 & l^7/7
    \end{pmatrix}
    \notag\\ 
    \bm{C}_2&=
    \begin{pmatrix}
        0 & 0 & 0 & 0 \\
        0 & 0 & 0 & 0 \\
        0 & 0 & 4l & 6l^2 \\
        0 & 0 & 6l^2 & 12l^3
    \end{pmatrix}
    \notag\\ 
    \bm{C}_3&=
    \begin{pmatrix}
        0 & 0 & 0 & 0 \\
        0 & 0 & 0 & 0 \\
        2l & l^2 & 2l^3/3 & l^4/2 \\
        3l^2 & 2l^3 & 3l^4/2 & 6l^5/5
    \end{pmatrix}.
\end{align}
Thus, $L_{\mathrm{H}}$ can be analytically calculated from the output $\hat{\bm{U}}$. Therefore, a difference approximation or numerical integration is not necessary to evaluate $L_{\mathrm{H}}$.

Finally, the loss function $L$ in the proposed method is defined as the weighted sum of the data loss function $L_{\mathrm{D}}$ and the Helmholtz equation loss function $L_{\mathrm{H}}$:
\begin{align}\label{eq:loss}
    L=L_{\mathrm{D}}+\lambda L_{\mathrm{H}},
\end{align}
where $\lambda$ is a balancing parameter. When $\lambda$ is set to $0$, the proposed method is equivalent to the baseline method in Sect.~\ref{sec:baseline}.

\section{Numerical experiments}\label{sec:exp}
Numerical experiments were conducted to evaluate the effectiveness of the learning-based framework and the proposed Helmholtz-equation loss function. The compared methods are the kernel interpolation method~\cite{Ueno:IWAENC2018,Ueno:IEEE_SPL2018} (\textbf{Kernel}), the baseline method described in Sect.~\ref{sec:baseline} (\textbf{Baseline}), and the proposed method (\textbf{Proposed}). 


\subsection{Experimental conditions}

\begin{figure}[t]
\begin{center}
\includegraphics[width=8cm]{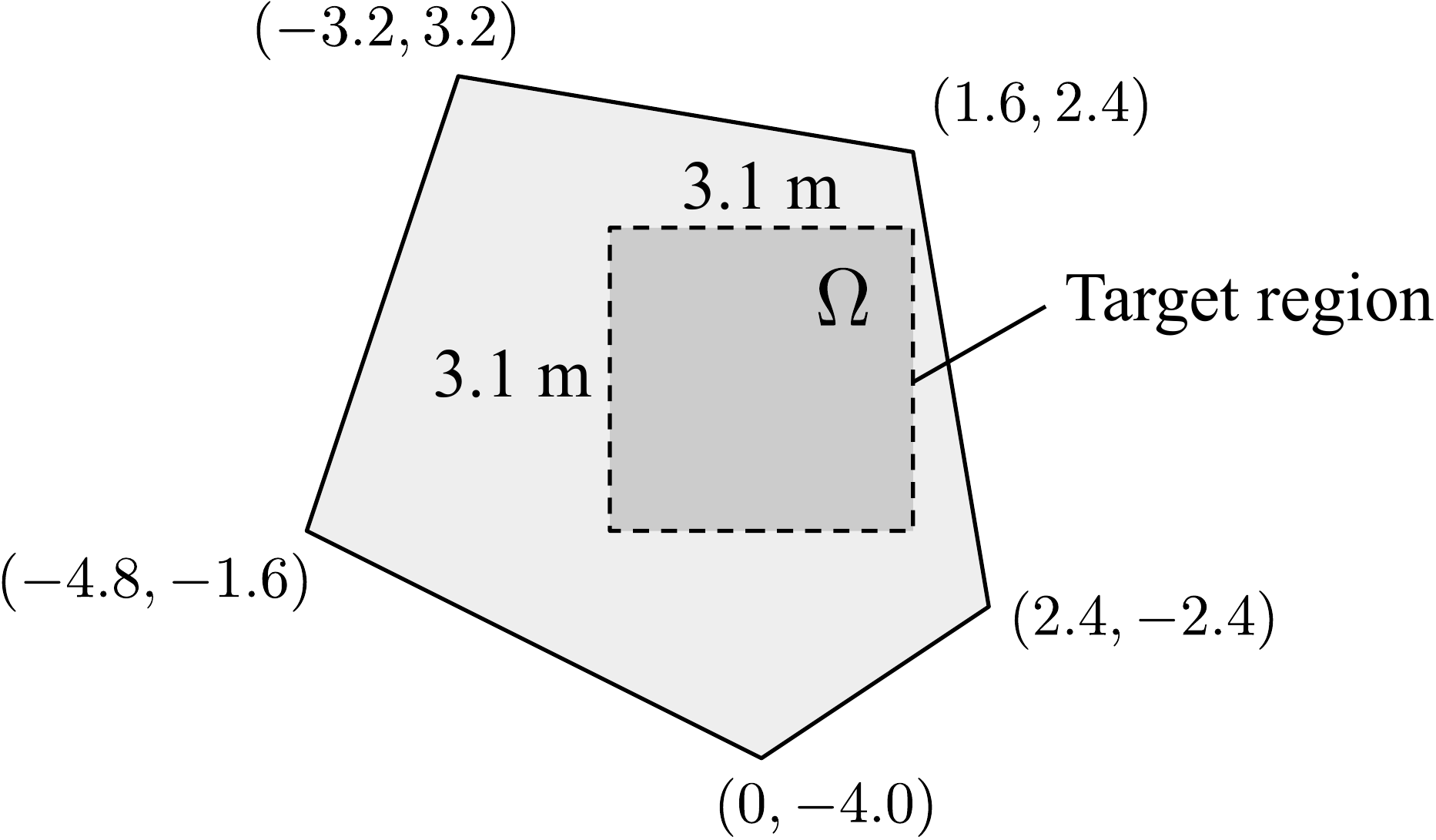}
\caption{Experimental setup. The sound field in the rectangular target region $\Omega$ is estimated. The positions of the vertices of the pentagonal room are indicated.}
\label{fig:pentaroom}
\end{center}
\end{figure}

The experiments were carried out under the assumption of a stationary sound field in a pentagonal room in a 2D space.
As shown in Fig.~\ref{fig:pentaroom}, the target rectangular region $\Omega$ was a square region of $3.1~\mathrm{m} \times 3.1~\mathrm{m}$ in size, whose center was at the coordinate origin. $I=J=32$ evaluation points were regularly set inside $\Omega$ at intervals of $0.1~\mathrm{m}$. The investigated frequency was $300~\mathrm{Hz}$. In \textbf{Kernel}, the kernel function of 0th-order Bessel function was used with the regularization parameter of $10^{-3}$. The details of the experimental conditions for \textbf{Baseline} and \textbf{Proposed} are described below. 

\subsubsection{Dataset}
A point source was placed at a randomly selected point outside $\Omega$. The sound field generated by the point source at a single frequency was simulated by the finite element method using FreeFEM++~\cite{MR3043640}. The sound absorption ratio on the wall was set at $0.25$. The sound speed was $340~\mathrm{m\slash s}$.
We generated the sound field data for $128\times 2$ point source positions. This dataset was divided into two equal parts: training and test data.
The number of observations $M$ were $5$, $10$, $15$, and $20$. The training was performed with randomly selected observation points from the evaluation points.

\subsubsection{Training}

The parameter $\lambda$ in \eqref{eq:loss} was set to $10^{-5}$ in \textbf{Proposed} on the basis of preliminary experimental results. 
For scale-independent learning, the input was standardized so that the value of each element was between $-1$ and $1$. Moreover, the phase was randomized in each epoch for phase-independent learning. 
CNN was trained using Adam~\cite{DBLP:journals/corr/KingmaB14}, and the learning rate was set at $0.01$.
For each dataset, training was repeated for $5000$ epochs.


\subsubsection{Evaluation measure}\label{subsec:evalidx}
For evaluation measures, normalized mean square error ($\mathrm{NMSE}$) and Helmholtz-equation error ($\mathrm{HE}$) are respectively defined as
\begin{align}
    &\mathrm{NMSE}=\dfrac{\sum_{(i,j)\in\mathcal{N}} \left| \hat{u}(\bm{r}_{i,j})-u(\bm{r}_{i,j}) \right|^2}{\sum_{(i,j)\in\mathcal{N}} \left| u(\bm{r}_{i,j}) \right|^2}\\
    &\mathrm{HE}=\frac{1}{S_\Omega}\int_{\Omega} \left|\left(\Delta +k^2\right)\hat{u}(x,y) \right|^2\mathrm{d}x\mathrm{d}y,
\end{align}
In \textbf{Baseline}, $\mathrm{HE}$ was evaluated for the function interpolated by bicubic spline interpolation from the CNN output, where $\hat{u}_x^{p}$ at $\mathcal{N}_x$, $\hat{u}_y^{p}$ at $\mathcal{N}_y$, and $\hat{u}_{x,y}^{p}$ at $\mathcal{N}_{x,y}$ were assumed to be $0$. Training was conducted 5 times for each number of observations $M$ in \textbf{Baseline} and \textbf{Proposed}.




\subsection{Results and discussion}

\begin{table}[t]
  \caption{Mean and standard deviation of $\mathrm{NMSE}$ in dB}
  \centering
  \begin{tabular}{c|ccc}
   \toprule
    $M$   & \textbf{Kernel} & \textbf{Baseline} & \textbf{Proposed} \\
   \midrule
     $5$ & $-1.36\pm0.69$ & $-2.44\pm1.66$ & $-2.44\pm1.65$ \\
     $10$ & $-3.21\pm1.11$ & $-4.87\pm2.34$ & $-4.88\pm2.33$ \\
     $15$ & $-5.99\pm1.66$ & $-6.58\pm2.70$ & $-6.61\pm2.68$ \\
     $20$ & $-11.02\pm2.64$ & $-7.87\pm2.96$ & $-7.92\pm2.92$ \\
  \bottomrule
    \end{tabular}\label{table:NMSEmean}
\end{table}

\begin{table}[t]
  \caption{Mean and standard deviation of $\mathrm{HE}$ on a logarithmic scale}
  \centering
  \vspace{-10pt}   
  


  \begin{tabular}{c|ccc}
  \toprule
    $M$   & \textbf{Kernel} & \textbf{Baseline} & \textbf{Proposed} \\
  \midrule
     $5$ & -- & $2.79\pm0.17$ & $2.22\pm0.17$ \\
     $10$ & -- & $2.95\pm0.17$ & $2.35\pm0.18$ \\
     $15$ & -- & $2.87\pm0.17$ & $2.22\pm0.19$ \\
     $20$ & -- & $2.75\pm0.16$ & $2.06\pm0.20$ \\
  \bottomrule
    \end{tabular}\label{table:HEmean}

\end{table}

\begin{figure}[t]
\begin{center}
\subfloat[Ground truth]{
    \includegraphics[height=3cm]{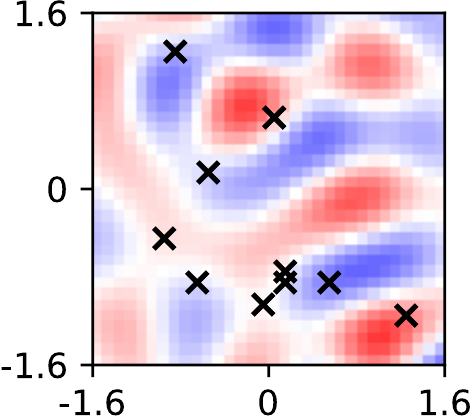}
\hspace{6pt}
    }
\subfloat[\textbf{Kernel}]{
\hspace{10pt}
    \includegraphics[height=3cm]{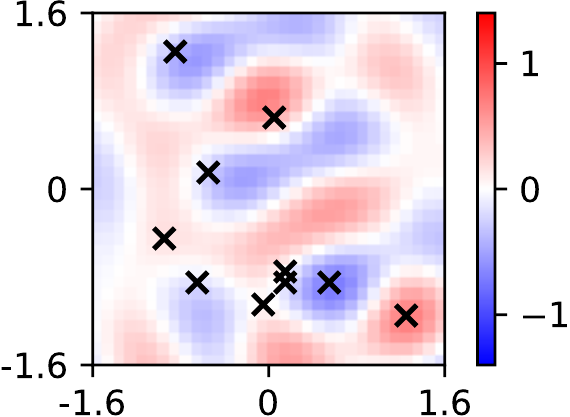}
    }
\par\medskip
\subfloat[\textbf{Baseline}]{
    \includegraphics[height=3cm]{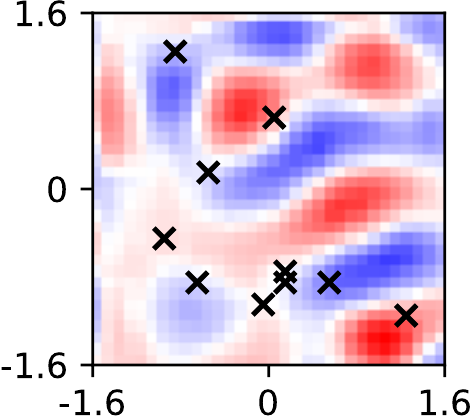}
    \hspace{6pt}
    }
\subfloat[\textbf{Proposed}]{
\hspace{10pt}
    \includegraphics[height=3cm]{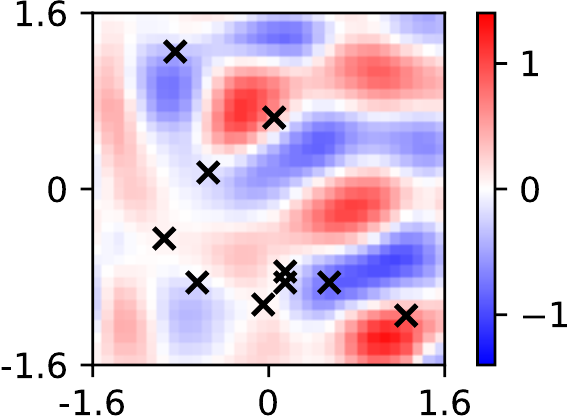}
    }
\caption{
Estimated pressure distribution when $M=10$. Ticks shows the scale of the target region in $\mathrm{m}$. Crosses indicate the observation points. NMSEs of \textbf{Kernel}, \textbf{Baseline}, and \textbf{Proposed} were $-4.95$, $-7.86$, and $-7.89$~dB, respectively. HEs on a logarithmic scale of \textbf{Baseline} and \textbf{Proposed} were $3.04$ and $2.37$, respectively.}
\label{fig:penta_output_visualize}
\end{center}
\end{figure}

The mean and standard deviation of $\mathrm{NMSE}$ in dB are shown in Table~\ref{table:NMSEmean}.
The learning-based methods, \textbf{Proposed} and \textbf{Baseline}, indicated improvements in $\mathrm{NMSE}$ for $M=5$, $10$, and $15$, compared with \textbf{Kernel}. The difference in $\mathrm{NMSE}$ between \textbf{Proposed} and \textbf{Baseline} was small. Therefore, the learning-based approach is effective when the number of sensors is small.

Fig.~\ref{fig:penta_output_visualize} shows an example of ground truth and estimated pressure distribution when $M=10$. Although the $\mathrm{NMSE}$ of \textbf{Baseline} is relatively small, the pressure distribution includes unnatural variations, compared with that of \textbf{Proposed}. These variations originate from the deviation of the estimate from the Helmholtz equation.

Table~\ref{table:HEmean} shows the mean and standard deviation of $\mathrm{HE}$ on a logarithmic scale.
Note that $\mathrm{HE}$ of \textbf{Kernel} is not indicated because the estimate of \textbf{Kernel} exactly satisfies the Helmholtz equation.
As indicated by $\mathrm{HE}$ in Table~\ref{table:HEmean}, the estimate of \textbf{Proposed} was less deviated from solutions of the Helmholtz equation, compared to that of \textbf{Baseline}. 


\section{Conclusion}
We proposed a learning-based sound field estimation method based on PICNN using bicubic spline interpolation. Although current learning-based methods enable us to estimate an acoustic field from sparse measurements, its solution can be physically infeasible because physical properties are not taken into consideration. We considered applying PICNN to the sound field estimation problem. However, direct application of PICNN can lead to errors because the CNN output is spatially discretized values of pressure distribution. We incorporate bicubic spline interpolation in the PICNN framework to evaluate the Helmholtz-equation loss function without discretization. In numerical experiments, the learning-based methods improved the estimation accuracy when the number of measurements is small. The estimation accuracy of the proposed method was comparable to that of the CNN-based method, but the deviation from the Helmholtz equation remained small, which means that a physically feasible solution can be obtained by the proposed method.

\section{Acknowledgements}
This work was supported by JSPS KAKENHI Grant Number 22H03608, JST FOREST Program (Grant Number JPMJFR216M, Japan), and OKI Electric Industry Co., Ltd.


\bibliographystyle{IEEEbib_mod}
\bibliography{str_def_abrv,refs}

\end{document}